Zr and Sn substituted $(Na_{0.5}Bi_{0.5})TiO_3$ –based solid solutions

V. M. Ishchuk[a], L. G. Gusakova[a], N.G. Kisel'[a], D.V. Kuzenko[a], N.A. Spiridonov[a], V. L. Sobolev[b]

[a]Science & Technology Center "Reactivelectron" of the National Academy of Sciences of Ukraine, Donetsk, 83049, Ukraine
[b]Deaprtment of Physics, South Dakota School of Mines and Technology, Rapid City, SD 57701, U.S.A.

**Abstract**

Formation of phases during the solid state synthesis of the $[(Na_{0.5}Bi_{0.5})_{0.80}Ba_{0.20}](Ti_{1-y}B_y)O_3$ system of solid solutions, with substitutions of zirconium and tin ions for titanium ones, has been investigated. It is demonstrated that the synthesis is a multi-step process which is accompanied by the formation of a number of intermediate phases (depending on the composition of the solid solution and the temperature of synthesis). Single phase solid solutions have been manufactured when the sintering temperature was increased to 1000 – 1100 °C.
   An increase of the concentration of the substituting ions results in a linear increase in the size of the crystal cell. As a consequence the reduction of the tolerance factor and an increase of the stability of the antiferroelectric phase with respect to the ferroelectric one take place.

## 1. Introduction

Lead zirconate-titanate, lead magnesium niobate–lead titanate, and lead zinc niobate-lead litanate solid solutions are materials with the largest values of piezoelectric parameters known at present. The lead zirconate-titanate (PZT) and PZT-based solid solutions are widely used as commercial ceramic materials. The necessity to develop lead-free substances with high values of piezoelectric parameters has stimulated intensive studies of these materials. The $(Bi_{0.5}Na_{0.5})TiO_3$-based piezoelectric materials are among them. The solid solutions $(1–x)(Bi_{0.5}Na_{0.5})TiO_3 – xBaTiO_3$ are the most investigated due to their promising piezoelectric behavior credited to the presence of the morphotropic phase boundary in compounds with x = 0.05–0.07 (similarly to PZT [1, 2] with high piezoelectric properties). However these solid solutions possess a number of disadvantages, such as a low temperature of destruction of the polar ferroelectric state (of the order of 130 – 150 °C) in the compounds from the morphotropic boundary region of the diagram of phase states of $(1–x)(Bi_{0.5}Na_{0.5})TiO_3 – xBaTiO_3$ and low values of the piezoelectric modulus $d_{33}$ (160 – 180 $pC/N$).

In the majority of published studies the investigated BNT-based solid solutions were manufactured by substitutions in the A-site of the perovskite crystal lattice. In all cases such substitutions led to the decrease of the temperature of the ferroelectric-antiferroelectric phase transition relative to the prime BNT. Furthermore, the BNT-based solid solutions have a large value of the coercive field which hinders their polarization process. All these factors do not allow BNT-based solid solutions to become an effective replacement for the PZT-based solid solutions.

However, it is well known that the large values of piezoelectric parameters in lead-containing substances are reached as consequence of the substitution of ions in the B-site. The solid solutions $PbZrO_3 – PbTiO_3$, $Pb(Mg_{1/3}Nb_{2/3})O_3 – PbTiO_3$, $Pb(Zn_{1/3}Nb_{2/3})O_3 – PbTiO_3$ are the good examples of this statement.

The aim of this study is to investigate the influence of the ion substitutions in the B-sites of the crystal lattice of the BNT-based solid solutions on the relative stability of the ferroelectric and antiferroelectric states and to examine the factors that promote the expansion of the temperature range of the existence of the ferroelectric state. We present the results of investigation of the effect of the substitution of zirconium and tin for titanium on the processes of formation of the solid solutions and the phase transitions in the synthesized substances.

## 2. Experimental methods

We have used a traditional method of solid state synthesis for manufacturing of the BNT-based solid solution. Reagent grade oxides and carbonates of corresponding metals have been used as starting materials which have been mixed in the appropriate stoichiometry (except for $Bi_2O_3$, that has been taken in excess of 0.5 Wt%) by ball milling during 12 hours. The mixed powders have been calcined at different temperatures in the interval $700 − 1100\,°C$ for $2 − 20$ hours.

The phase analysis was carried out by DRON-3 X-ray diffractometer in the Bragg-Brentano geometry, using Cu$K\alpha$-radiation ($\lambda = 1.5418$ Å), Ni-filter for incident beam, and graphite monochromator in the diffraction beam (the angular range was $20° \leq 2\theta \leq 90°$, the scan step was $0.02°$, the accumulation time at each point was 1 s). The intermediate phases were identified using the ASTM-library.

The synthesized powders were axially pressed into disks with a diameter of 12 mm and a thickness of 1 mm. After that they were sintered at $1150 – 1200\,°C$ to manufacture samples for dielectric measurements.

A fire-on silver paste was used for electrodes. Temperature dependencies of dielectric parameters were measured at 1 kHz by QuadTech 7600 LCR meter. *D–E* hysteresis loops were observed by a standard Sawyer-Tower circuit at $10^{-2}$ Hz.

## 3. Results of solid state reactions studies

In the PZT solid solutions one of the components ($PbTiO_3$) is a ferroelectric and the other one ($PbZrO_3$) is an antiferroelectric. The increase of the content of zirconium which substitutes the titanium in the crystal lattice leads to a stabilization of the nonpiezoelectric antiferroelectric phase [3, 4]. There are very few publications (see [5-8], for example) containing results on substitution of zirconium for titanium in the NBT-based solid solutions. However, it has been expected that the substitution of zirconium would also lead to the stabilization of the antiferroelectric state in these compounds. Therefore, we have chosen the solid solution $[(Na_{0.5}Bi_{0.5})_{0.8}Ba_{0.2}]TiO_3$ as a base one for our investigations of the influence of substitutions of titanium by zirconium The position of the $[(Na_{0.5}Bi_{0.5})_{0.8}Ba_{0.2}]TiO_3$ compound in the diagram of phase states of the system $[(Na_{0.5}Bi_{0.5})]TiO_3 - BaTiO_3$ is located in the region of the ferroelectric state stability [9, 10].

### 3.1. $[(Na_{0.5}Bi_{0.5})_{1-x}Ba_x]TiO_3$ solid solutions

Here we present results of the investigation of the $[(Na_{0.5}Bi_{0.5})_{1-x}Ba_x]TiO_3$ solid solutions for comparison with other result that will follow. The introduction of barium ions into the crystal lattice led to the more complicated process of formation of phases in the system of initial components compared to the same process of manufacturing of the pure $(Na_{0.5}Bi_{0.5})TiO_3$ (see Fig. 1a and b). When the mixture of the starting components with the $x = 0.1$ Ba-content was annealed at 700 °C during 6 hours the reaction product contained a large amount of the $(Na_{0.5}Bi_{0.5})TiO_3$-based solid solution along with the initial oxides and carbonates, and the $BaTiO_3$-based solid solution in a small amount. The reaction products also contained the following compounds: bismuth titanate $Bi_{12}TiO_{20}$, sodium bismuth titanate $NaBiTi_6O_{14}$, barium titanate $BaTiO_3$, bismuth $BaBi_4TiO_{15}$ and $BaBiO_{2.77}$ (all in very small amount).

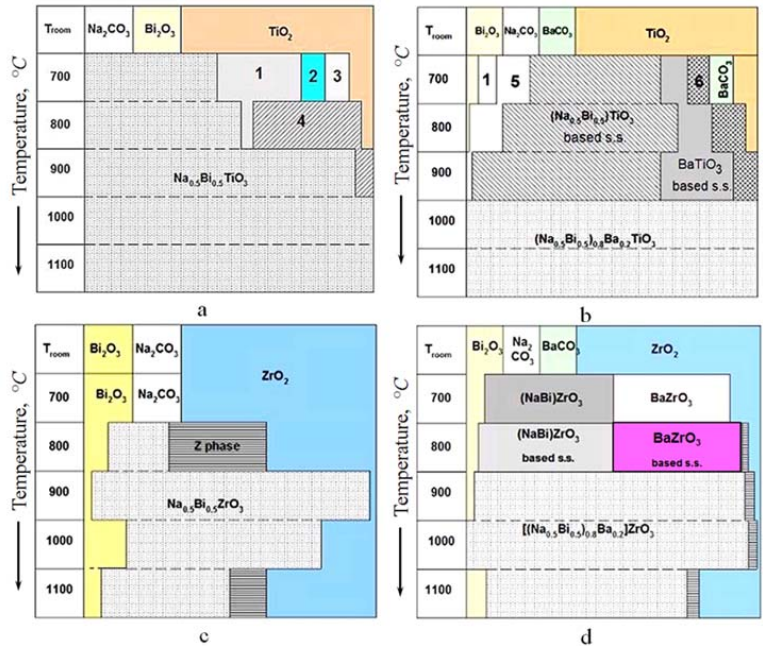

Fig.1. Schematic representation of the formation of phases during the solid state synthesis of $(Na_{0.5}Bi_{0.5})TiO_3$ (a); $[(Na_{0.5}Bi_{0.5})_{0.8}Ba_{0.2}]TiO_3$ (b); $(Na_{0.5}Bi_{0.5})ZrO_3$ (c); and $[(Na_{0.5}Bi_{0.5})_{0.8}Ba_{0.2}]ZrO_3$ (d).
Intermediate phases: 1 – $Bi_{12}TiO_{20}$, 2 – $Na_2Ti_6O_{13}$, 3 – $Na_2Ti_3O_7$, 4 – $Bi_4Ti_3O_{12}$, 5 – $NaBiTi_6O_{14}$, 6 – $BaBi_4Ti_4O_{15}$.

When the synthesis temperature increased to 800 °C with the same 6 hours duration of synthesis the content of bismuth titanate $Bi_{12}TiO_{20}$ and sodium bismuth titanate $NaBiTi_6O_{14}$ decreased, and compounds $BaTiO_3$ and $BaBiO_{2.77}$ disappeared completely. Almost a single-phase solid solution was formed at sintering at 900 °C.

The sequence of phases during the synthesis of the $[(Na_{0.5}Bi_{0.5})_{1-x}Ba_x]TiO_3$ solid solutions with increased Ba-content to x = 0.2 at low sintering temperature (700 − 800 °C) was analogous to the sequence of the phase formation for the system with $x$ = 0.1. But the complete formation of a single-phase solid solution in this case required a higher temperature of about 1000 °C (see Fig. 1b).

### 3.2. $[(Na_{0.5}Bi_{0.5})_{1-x}Ba_x]ZrO_3$ solid solutions

We consider the synthesis of the sodium bismuth zirconate compound without Ba first. The formation of $(Na_{0.5}Bi_{0.5})ZrO_3$ (NBZ), with slightly distorted perovskite lattice from the $0.25·Na_2CO_3$ + $(0.25 + 0.5\ Wt\%)·Bi_2O_3 + ZrO_2$ mixture of reagents began at 800 °C. The reaction product contained the new Z-phase. Based on the X-ray diffraction data we assumed that Z-phase represents $(Na_{0.5}Bi_{0.5})ZrO_3$ with excess of Bi and deficiency of Zr in the crystal lattice. After the annealing at 900 °C for nearly 3 hours took place the NBZ was the primary phase; $ZrO_2$ and $Bi_2O_3$ were present in a small amount and the Z-phase was not observed.

An increase of the annealing temperature up to 1000 °C and/or an increase of the sintering time led to a decrease of the content of the sodium-bismuth zirconate in the synthesis products. The results of chemical analysis showed that this was due to the volatility of the sodium oxide (the loss of bismuth was much lower). The schematic of the process of formation of the sodium bismuth zirconate is presented in Fig. 1c.

Substitutions in the *A* sublattice of $(Na_{0.5}Bi_{0.5})$ by barium led to a few changes the nature of phase transformations during the solid-state reactions in comparison with NBZ. The reaction in the mixture of the starting components with the formula $[(Na_{0.5}Bi_{0.5})_{0.90}Ba_{0.10}]ZrO_3$ started at a lower temperature of 700 °C, and was accompanied by the formation of $(Na_{0.5}Bi_{0.5})ZrO_3$, $(Na_{0.5}Bi_{0.5})ZrO_3$-based solid solutions, and $BaZrO_3$. This last result is interesting and surprising because the formation of barium zirconate in the conventional solid-state synthesis takes place at the temperatures in the range 1200 – 1300 °C or higher. Moreover, the reaction mixture contained the starting components $ZrO_2$ and $Bi_2O_3$ in rather large amounts.

The share of the NBZ-based solid solution increased and the share of the $BaZrO_3$-based one decreased when the sintering temperature increased to 800 °C. The main product of the reaction, namely, $[(Na_{0.5}Bi_{0.5})_{0.90}Ba_{0.10}]ZrO_3$ with small distortion of perovskite crystal lattice was achieved as a result of annealing at 900 °C during 10 hours. A small amount of $ZrO_2$ and the Z-phase were also present. The barium zirconate $BaZrO_3$ was absent. The time of not less than 20 hours was necessary to complete formation of the single-phase product at this temperature.

An increase of the sintering temperature to 1100 °C led to a reduction of the share of $[(Na_{0.5}Bi_{0.5})_{0.90}Ba_{0.10}]ZrO_3$ in the reaction mixture. The $Bi_2O_3$ and $ZrO_2$ basic components appeared along with this. Their appearance indicated the decomposition of the solid solution.

The increase of the barium content led to the complication of the phase formation process at low temperatures (up to 900 °C), however, it gave a higher yield of the final product after the synthesis at higher temperatures (see Fig. 1d). The mixture of components, corresponding to the composition $[(Na_{0.5}Bi_{0.5})_{0.80}Ba_{0.20}]ZrO_3$ contained $(Na_{0.5}Bi_{0.5})ZrO_3$-based solid solutions with the tetragonal perovskite structure ($a$ = 4,08 Å, $c$ = 4,101 Å), $BaZrO_3$-based solid solutions with the cubic structure ($a$ = 4.186 Å) and small amounts of the basic oxides $Bi_2O_3$ and $ZrO_2$ after the heat treatment at 700 °C. Further heat treatment in the temperature range 900-1000 °C during 3 hours was accompanied by the formation of the $[(Na_{0.5}Bi_{0.5})_{0.80}Ba_{0.20}]ZrO_3$ solid solution with cubic perovskite structure ($a$ = 4.125 Å). Oxides $Bi_2O_3$, $ZrO_2$ and the Z-phase were present in small quantities after annealing at 900 °C. $ZrO_2$ and Z-phase were observed in small quantities after the annealing at 1000 °C. However, they disappeared when the synthesis time increased.

An increase of the annealing temperature to 1100 °C or an increase of the sintering time to 24 hours at 1000 °C led to a decrease of the $[(Na_{0.5}Bi_{0.5})_{0.80}Ba_{0.20}]ZrO_3$ share, and to an increase of the initial oxides $Bi_2O_3$ and $ZrO_2$ content. The Z-phase was present in a very small amount. The perovskite lattice parameter, $a$, increased and became to be equal to 4.131 Å. It should be noted that an increase of the barium content in the solid solutions led to the situation that the share of the final product $(Na_{0.5}Bi_{0.5})_{1-x}Ba_xZrO_3$ increased at the high-temperature annealing and in the presence of the decomposition of the solid solution.

Additional low-temperature annealing at 750 °C during 6 hours led to the disappearance of the Z-phase. The contents of the $Bi_2O_3$ and $ZrO_2$ oxides and the $[(Na_{0.5}Bi_{0.5})_{0.80}Ba_{0.20}]ZrO_3$ solid solution remained almost unchanged. The lattice parameter of the cubic lattice of the solid solution remained unchanged also.

In all cases when the synthesis of solid solutions was carried out at the temperatures below 700 °C, the following intermediate compounds were present in the final mixture: $Bi_4Ti_3O_{12}$, $Na_2Ti_6O_{13}$, $Na_2Ti_3O_7$, $Bi_2Ti_2O_7$, $Ba_2TiO_4$, $BaTi_4O_9$, $NaBiTi_6O_{14}$. After the following annealing at the higher temperatures, they were not observed.

### 3.3. $[(Na_{0.5}Bi_{0.5})_{0.80}Ba_{0.20}](Ti_{1-y}Zr_y)O_3$ solid solutions

To analyze the formation of solid solutions containing the titanium and zirconium ions in the B-sites of the crystal lattice we have considered the $[(Na_{0.5}Bi_{0.5})_{0.80}Ba_{0.20}](Ti_{0.90}Zr_{0.10})O_3$ (with $y = 0.10$) as an example. After the annealing at 700 °C during 6 hours the reaction mixture contained approximately equal amounts of two solid solutions on the basis of $[(Na_{0.5}Bi_{0.5}),Ba](Ti_{0.5}Zr_{0.5})O_3$ with a cubic crystal structure and the lattice parameters $a = 3.90$ Å and $a = 4.02$ Å. Apparently, the first solid solution contained less zirconium and the second solid solution contained more barium than the first one. The compounds $Bi_{12}TiO_{20}$, $Bi_2Ti_2O_7$, and the Z-phase (defined in the previous section) were present. The $[(Na_{0.5}Bi_{0.5})_{0.80}Ba_{0.20}](Ti_{0.90}Zr_{0.10})O_3$ compound was also present (see Fig.2).

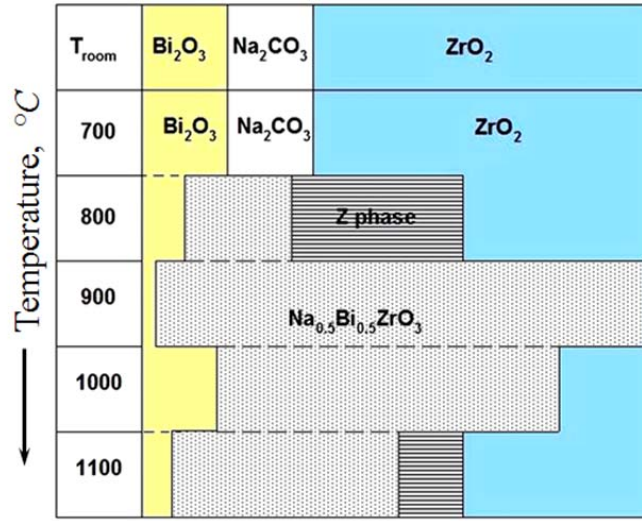

Fig.2. Schematic of the phases formation during the solid state synthesis of the $[(Na_{0.5}Bi_{0.5})_{0.80}Ba_{0.20}](Ti_{1-y}Zr_y)O_3$ solid solutions for y = 0.1.
Intermediate phases: 1 – $Bi_{12}TiO_{20}$, 2 – $Ba_2Ti_2O_7$, 3 – $Ba_2TiO_4$, 4 – $BaTi_4O_9$.

The $ZrO_2$ and Z-phase disappeared in the reaction mixture when the synthesis temperature was increased to 900 °C. Intermediate compound $Bi_{12}TiO_{20}$ at this sintering temperature was present in very minor amounts and was preserved even at a temperature of 1000 °C. The reaction products were a mixture of solid solutions with lattice parameters $a = 3.94$ Å (main product) and $a = 4.00$ Å after synthesis at 1000 °C during 6 hours. The share of the second solid solution decreased with an increase of the temperature and time of exposure. A single-phase $[(Na_{0.5}Bi_{0.5})_{0.80}Ba_{0.20}](Ti_{0.90}Zr_{0.10})O_3$ solid solution with the cubic crystal structure and the lattice parameter $a = 3.960$ Å was formed at 1100 °C.

The sequence of phases formed during synthesis process was somewhat more complicated when the content of zirconium increased to 20%. Two predominant solid solutions (the same as in the compound with y = 0.1) were present after calcination at 700 °C. However, barium carbonate $BaCO_3$ and a new intermediate phase with the perovskite crystal structure ($a$ = 4.35 Å) were also detected in addition to the original components. There was also some amount of $Bi_{12}TiO_{20}$ (as in the case of $y$ = 0.10).

Appreciable amounts of $ZrO_2$, as well as $Bi_{12}TiO_{20}$ still remained in the reaction products when the calcination temperature was increased to 800 °C. Two solid solutions $[(Na_{0.5}Bi_{0.5})_{0.80}Ba_{0.20}](Ti_{0.90}Zr_{0.10})O_3$ (with $a$ = 3,94 Å) and $[(Na_{0.5}Bi_{0.5})Ba](Ti_{0.5}Zr_{0.5})O_3$-based solid solution (with $a$ = 4.02 Å) were in a predominant amount. The reaction products were a mixture of solid solutions with the lattice parameters $a$ = 3.95 Å and $a$ = 4.02 Å after 6 hours of synthesis at 900 and 1000 °C. A single-phase solid solution with the lattice parameter $a$ = 3.974 Å was formed at 1100 °C. The decomposition of the solid solution at higher temperatures did not take place.

The increase in the content of zirconium up to 30% did not lead to significant changes in the pattern of the solid solutions formation.

Concluding the discussion of the results of study of the phase formation in BNT-based solid solutions containing zirconium, we have to discuss the decomposition of solid solutions at high temperatures. In the case of the solid solutions on the basis of $(Na_{0.5}Bi_{0.5})ZrO_3$ without titanium this decomposition took place. In the other case of the $[(Na_{0.5}Bi_{0.5})_{0.80}Ba_{0.20}](Ti_{1-y}Zr_y)O_3$ solid solutions containing titanium the decomposition did not occur. In the first case, the synthesis of the solid solution did not end with 100% yield of the finished product. It should be emphasized also that the introduction of the Ba-ions into the crystal structure of $(Na_{0.5}Bi_{0.5})ZrO_3$ reduced the degree of the solid solution decomposition at high temperatures. As noted above, the decomposition of the solid solution was accompanied primarily by the volatilization of the sodium oxide.

The transition from $(Na_{0.5}Bi_{0.5})TiO_3$ to $(Na_{0.5}Bi_{0.5})ZrO_3$ in the process of ion substitution is accompanied by an increase in the size of the crystal lattice due to the significant difference in the ionic radii of the $Ti^{4+}$ and $Zr^{4+}$ ions (0.605 and 0.72 Å, respectively[11]). Two major factors one must take into account. The ionic radius of $Na^+$ is less than the ionic radius of $Bi^{3+}$ into 12-coordination position of the perovskite crystal structure. Only one electron participates in the formation of bonding in the crystal lattice in the case of sodium, while in the case of bismuth there are three electrons for bonding. For these reasons the increase of the interionic distance leads to a more rapid decrease in bonding energy for sodium than for bismuth. Therefore, the sodium volatilization and the solid solution decomposition at high temperatures have occurred in the case of zirconium compounds.

In the $[(Na_{0.5}Bi_{0.5})_{0.80}Ba_{0.20}](Ti_{1-y}Zr_y)O_3$ solid solutions with substitution of titanium by zirconium (up to 30%) the change in the lattice parameters is much less (than for the solid solutions considered earlier), so the sodium is coupled to crystal structure stronger, and the high-temperature decomposition of the solid solution does not take place. As demonstrated above the formation of the solid solution occurred at higher temperatures than in the absence of zirconium, and a solid state synthesis process was more complicated.

Thus, the synthesis of the $[(Na_{0.5}Bi_{0.5})_{0.80}Ba_{0.20}](Ti_{1-y}Zr_y)O_3$ system of solid solutions is a multistage process which is accompanied (depending on the composition of the solid solution) by the formation of a number of the following intermediate phases: $Bi_{12}TiO_{20}$, $Bi_4Ti_3O_{12}$, $Na_2Ti_6O_{13}$, $Na_2Ti_3O_7$, $Bi_2Ti_2O_7$, $Ba_2TiO_4$, $BaBi_4Ti_4O_{15}$, $BaTi_4O_9$, $NaBiTi_6O_{14}$, $BaZrO_3$. Barium zirconate $BaZrO_3$ is detected in the reaction products after annealing at 700 °C and is present in the reaction mixture when the annealing temperature increases to 850 °C. In the absence of titanium the high-temperature annealing leads to decomposition of the synthesized solid solutions. Barium activates the synthesis process, reducing the onset temperature of the solid-state reaction in the $[(Na_{0.5}Bi_{0.5})_{1-x}Ba_x]ZrO_3$ system compared to the $(Na_{0.5}Bi_{0.5})ZrO_3$. High temperature annealing leads to the decomposition of solid solutions, if the $B$-sites of the crystal lattice contain Zr ions only. In the solid solution containing titanium the decomposition has not been observed during the high-temperature annealing.

Ceramic samples for further research were manufactured by the sintering of powders at 1200 °C for 6 hours. X-ray diffraction patterns of the $[(Na_{0.5}Bi_{0.5})_{0.80}Ba_{0.20}](Ti_{1-y}Zr_y)O_3$ solid solutions with $0.00 \leq y \leq 0.30$ are shown in Fig. 3. All samples were single-phase.

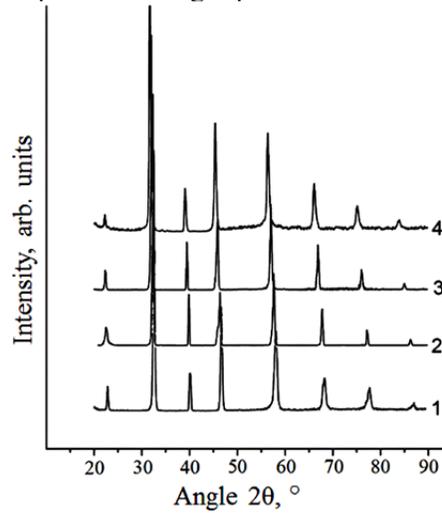

Fig.3. X-ray patterns for the $[(Na_{0.5}Bi_{0.5})_{0.80}Ba_{0.20}](Ti_{1-y}Zr_y)O_3$ solid solutions. Zr-content (y): 1 – 0.00, 2 – 0.05, 3 – 0.10, 4 – 0.30.

3.4. $[(Na_{0.5}Bi_{0.5})_{0.8}Ba_{0.2}](Ti_{1-y}Sn_y)O_3$-based solid solutions

The synthesis process of the $(Na_{0.5}Bi_{0.5})TiO_3$ based solid solutions with tin substitutions for titanium is simpler than in the case of the zirconium substitutions because the ion radius of $Sn^{4+}$ is smaller than the one of $Zr^{4+}$.

The synthesis of the $[(Na_{0.5}Bi_{0.5})_{0.8}Ba_{0.2}](Ti_{1-y}Sn_y)O_3$-compouds with $y = 0.10$ and $y = 0.20$ with annealing at 800 °C was accompanied by the formation of the solid solution with the perovskite pseudocubic crystal structure with the lattice parameter $a = 3.93$ Å for $y = 0.1$ and $a = 3.953$ Å for $y = 0.2$, as well as by the formation of a solid solution with a fluorite-type structure ($a = 5.28$ Å). The predominant phase was the first solid solution. The amount of the solid solution with fluorite type structure increased with the increase of the Sn content. Another compound with the perovskite crystal structure and $a = 4.01$ Å was also present in a trace amount. Considering the difference in the sizes of titanium and tin ions, it can be assumed that the first of the two solid solutions with perovskite crystal structure have been formed on the basis of $[(Na_{0.5}Bi_{0.5})_{0.8}Ba_{0.2}]TiO_3$ with the substitution of titanium by tin, and the second one is formed on the basis of $[(Na_{0.5}Bi_{0.5})_{0.8}Ba_{0.2}]SnO_3$ with the substitution of tin by titanium.

The content of the phase with the fluorite structure in the reaction mixture decreased by approximately to 5% for $y = 0.1$ and by less than 10% for $y = 0.2$ when the annealing temperature was 900 °C. The main products of the reaction in both cases were the solid solutions with perovskite structure and the lattice parameters $a = 3.93$ Å for $y = 0.1$ and $a = 3.953$ Å for $y = 0.20$. A second solid solution with the perovskite structure was present in an amount of 5 - 7%. The minor amounts of the tin oxide $SnO_2$ (2-3%) were also present.

The complete formation of the single-phase solid solution depends on the Sn-content and occurs when the sintering temperature is in a range of 1000-1100 °C.

## 4. Influence of ion substitutions on the phase transitions

Temperature dependencies of the real and imaginary parts of the dielectric constant of obtained solid solutions were measured to study the influence of ion substitutions on the phase transitions. Fig. 4a shows the temperature dependencies of the real part of the dielectric constant, $\varepsilon'$, measured on ceramic samples of the $[(Na_{0.5}Bi_{0.5})_{0.80}Ba_{0.20}](Ti_{1-y}Zr_y)O_3$ solid solutions with different zirconium content. Two anomalies are present in these dependencies. The first of anomaly (the low temperature one) corresponds to the ferroelectric−antiferroelectric phase transition and the second one corresponds to the transition from the antiferroelectric (non-polar) to the paraelectric phase. The low temperature phase transition is accompanied by an anomaly in the temperature dependence of the imaginary part of the dielectric constant, $\varepsilon''$, (see Fig. 4b). As one can see the effect of zirconium on the position of the high-temperature phase transition is weak. The temperature, at which the low-temperature phase transition occurs, decreases significantly when the content of zirconium in the solid solution increases.

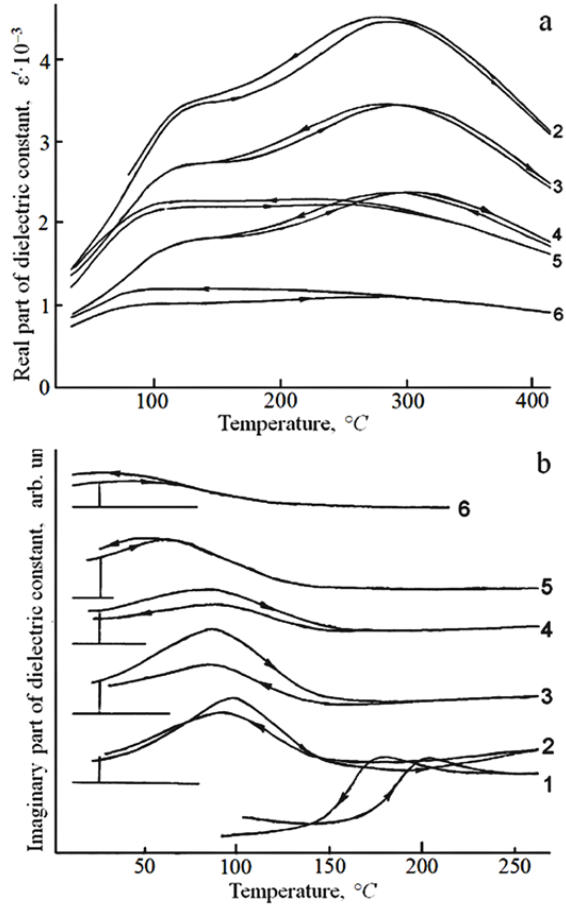

Fig.4. Temperature dependencies of the real (a) and imaginary (b) part of dielectric constant for the $(Na_{0.5}Bi_{0.5})_{0.80}Ba_{0.20}](Ti_{1-y}Zr_y)O_3$ solid solutions.
*Zr*-content, (*y*): 1 – 0.00, 2 – 0.025, 3 – 0.05, 4 – 0.10.

The study of the crystal structure of these solid solutions showed that an increase in zirconium content led to a linear increase in the size of the crystal unit cell. Fig.5 shows the dependencies of the crystal lattice parameter (calculated in pseudocubic approximation) on the content of zirconium. As we can see, the linear dependence takes place with good accuracy.

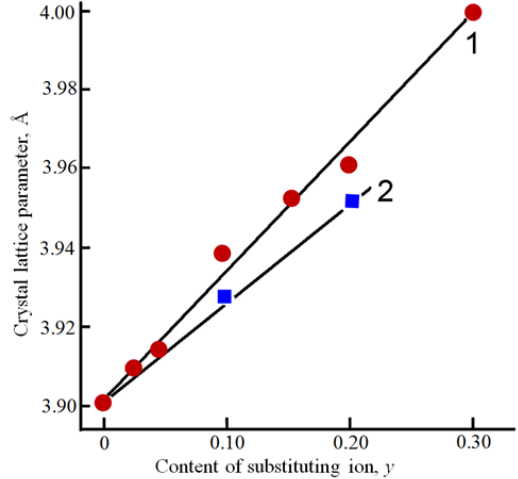

Fig.5. Dependencies of the crystal lattice parameter of the $[(Na_{0.5}Bi_{0.5})_{0.80}Ba_{0.20}]TiO_3$ based solid solution on Zr and Sn content.
Subsisting element: 1 – Zr, 2 – Sn.

Similar results were obtained for the solid solutions with substitutions of tin for titanium. The growths in the tin content led to a decrease of the temperature of the transition between the ferroelectric and antiferroelectric states and to a linear increase of the size of the crystal unit cell (see Fig.5). Since the size of the Sn ion is smaller than the Zr one, the rate of increase of the lattice parameter in the case of substitutions of Ti by Sn is smaller than in the case of substitutions by Zr. The temperature of the phase transition between the two dipole-ordered states (ferroelectric and antiferroelectric) decreased when the tin content increased (see Fig.6).

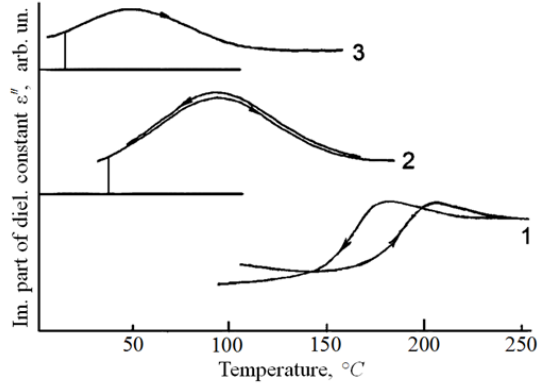

Fig.6. Imaginary part of dielectric constant vs. temperature for the $(Na_{0.5}Bi_{0.5})_{0.80}Ba_{0.20}](Ti_{1-y}Sn_y)O_3$ solid solutions.
Sn-content ($y$): 1 – 0.00, 2 – 0.10, 3 – 0.20.

Effect of the zirconium and tin ions on the relative stability of the ferroelectric and antiferroelectric states can be explained on the basis of size effect, as it has been done in relation to the stability of the same phases in the PZT-based solid solutions. As an estimation parameter tolerance factor $t = (R_A + R_O)/\sqrt{2}(R_B + R_O)$ is usually taken. In the PZT-based solid solutions the ferroelectric state is a stable state at large values of $t$ ($t > 0.9090$), for small values of $t$ ($t < 0.9080$) the stable state is the antiferroelectric one [12]. As it follows form this formula, an increase in the ionic radius of the ion in the $B$-site leads to a decrease of the $t$-factor. Just this effect has been manifested in the case of the substitution of titanium by zirconium and by tin in the $(Na_{0.5}Bi_{0.5})TiO_3$-based solid solutions.

## 5. Conclusions

Synthesis of the $(Na_{0.5}Bi_{0.5})TiO_3$ based solid solutions with substitution of Zr ions for Ti ions is a multistep process which is accompanied by the formation of a number of the following intermediate phases: $Bi_{12}TiO_{20}$; $Bi_4Ti_3O_{12}$; $Na_2Ti_6O_{13}$, $Na_2Ti_3O_7$, $Bi_2Ti_2O_7$, $Ba_2TiO_4$, $BaTi_4O_9$, $NaBiTi_6O_{14}$, and $BaZrO_3$.

Substitution of the Sn ions for Ti ones lead to a decrease of the number of intermediate compounds in comparison with the Zr substitution. An intermediate phase with the fluorite-type crystal structure appears during annealing at low temperatures in the case of Sn substitutions.

Single phase solid solutions have been obtained in all cases when the sintering temperatures were 1000 °C and above.

The substitutions of both zirconium and tin for titanium have led to a decrease of the tolerance factor and as a consequence to an increase of the stability of the antiferroelectric phase relative to the stability of the ferroelectric phase.